\documentclass[manuscript]{aastex}

\usepackage{hyperref}

\shorttitle{Nonlinear time-series analysis of Hyperion's lightcurves}
\shortauthors{Tarnopolski}

\begin{document}

\title{Nonlinear time-series analysis of Hyperion's lightcurves}

\author{M. Tarnopolski}
\affil{Astronomical Observatory of the Jagiellonian University\\ul. Orla 171, 30-244 Cracow, Poland}
\email{mariusz.tarnopolski@uj.edu.pl}

\begin{abstract}
Hyperion is a satellite of Saturn that was predicted to remain in a chaotic rotational state. This was confirmed to some extent by Voyager~2 and Cassini series of images and some ground-based photometric observations. The aim of this aticle is to explore conditions for potential observations to meet in order to estimate a maximal Lyapunov Exponent (mLE), which being positive is an indicator of chaos and allows to characterise it quantitatively. Lightcurves existing in literature as well as numerical simulations are examined using standard tools of theory of chaos. It is found that existing datasets are too short and undersampled to detect a positive mLE, although its presence is not rejected. Analysis of simulated lightcurves leads to an assertion that observations from one site should be performed over a year-long period to detect a positive mLE, if present, in a reliable way. Another approach would be to use 2---3 telescopes spread over the world to have observations distributed more uniformly. This may be achieved without disrupting other observational projects being conducted. The necessity of time-series to be stationary is highly stressed.
\end{abstract}

\keywords{chaos -- planets and satellites: individual: Hyperion}

	\section{Introduction}\label{sect1}

Saturn's seventh moon, Hyperion, was discovered in the XIX century by \citet{bond} and \citet{lassel}, but it took more than a century to obtain its images due to Voyager~2 \citep{smith} and Cassini \citep{thomas} missions. Its shape is highly elongated (360$\times$266$\times$205 km), making it the biggest known highly aspherical celestial body in the Solar System. \citet{wisdom} predicted Hyperion to remain in a chaotic rotational state due to its high oblateness and relatively high eccentricity, $e=0.1$. In dynamical system theory, a chaotic behaviour is recognised through a positive maximal Lyapunov Exponent (mLE), which describes the rate of divergence (or convergence in the negative case) of initially nearby phase-space trajectories. The Lyapunov spectrum is relatively easy to calculate in the case when the differential equations are known \citep{benettin,wolf,sandri,baker,ott}. On the other hand, there exist algorithms allowing to obtain an mLE from an experimental or observational time-series \citep{wolf,kantz94}, although they are to be used with carefullness, using at least a few hundred data points \citep{rosenstein, katsev} for nonlinear analysis. It is a hard task in astronomy to obtain long-term, well-sampled lightcurves. Although, despite this difficulty, it has been efficiently shown that pulsar spin-down rates exhibit chaotic dynamics \citep{seymour}: by re-sampling the original measurements, artificial time-series were produced, equivalent to the original ones, containing such a number of data points that the calculation of the correlation dimension and the mLE of the attractor, reconstructed {\it via} Takens time delay embedding method, was possible.

Hyperion's long-term observations were carried out twice in the post Voyager~2 era. In 1987, \citet{klav} (hereinafter, K89) performed photometric $R$ band observations over a timespan of more than 50 days, resulting in 38 high-quality data points. In 1999 and 2000, \citet{devyatkin} (hereinafter, D02) conducted $C$ (integral), $B$, $V$ and $R$ band observations. To the author's knowledge \citetext{Mel'nikov, priv.\ comm.} there were no other long-term observations that resulted in a lightcurve allowing to determine the rotational state of Hyperion. Although, shortly after the Cassini 2005 passage a ground-based $BVR$ photometry was conducted \citep{hicks}, resulting in 6 nights of $BVR$ measurements (and additional 3 nights of $R$ photometry alone) over a month-long period. Unfortunately, this data was greatly undersampled and period fitting procedures yielded several plausible solutions. 

The period determination in K89 was performed using the Phase Dispersion Minimization \citep{stellingwerf} and in D02 the Deeming method \citep{deeming} was used. Both authors noted that the periods found (6.6 d and 13.8 d in K89, and 10.8 d in D02) were statistically insignificant, leading to the conclusion that Hyperion's rotation is chaotic.

This work is focused on estimating the mLE based on existing observations and comparison of the results with the interpretation of their authors. Moreover, conditions that future observations should meet in order to reliably estimate the mLE are discussed. The paper is organized in the following manner. Section~\ref{sect2} describes the existing lightcurves used. Section~\ref{sect3} briefly explains the numerical methods used to calculate the mLE and presents their results, while in Section~\ref{sect4} numerical experiments on simulated lightcurves are performed. Section~\ref{sect5} is devoted to discussion and concluding remarks.

	\section{Datasets}\label{sect2}

\begin{figure}
\plotone{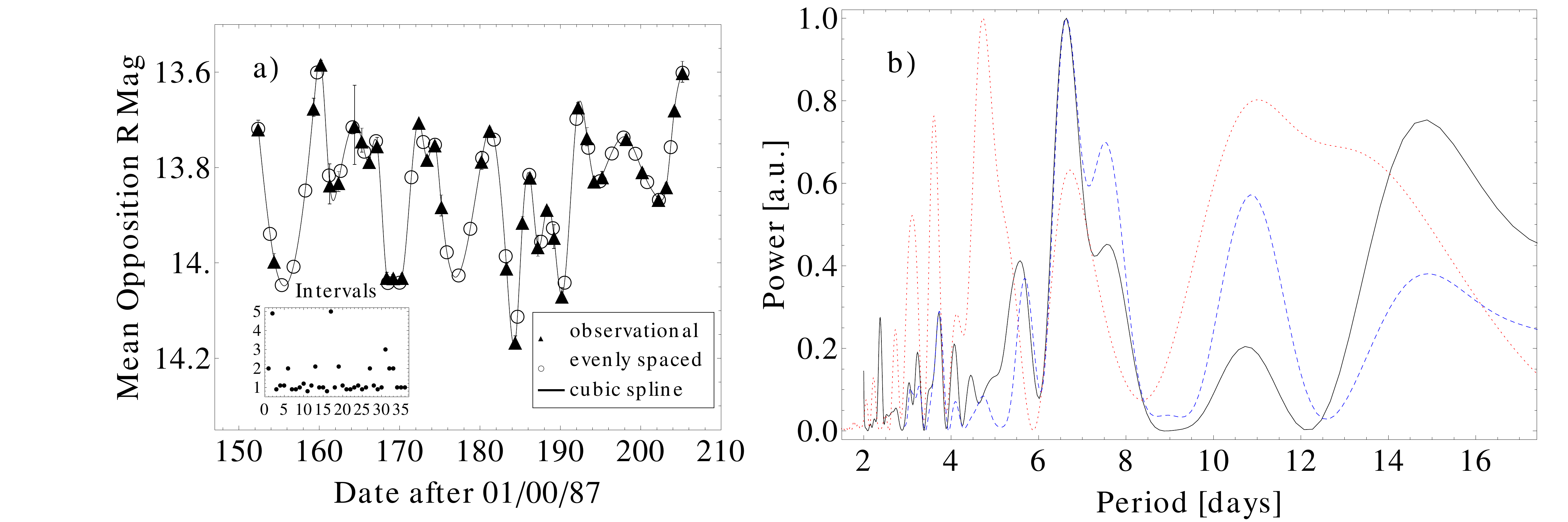}
\caption{a) The original K89 data (filled triangles) and the equally spaced re-sampled ones (open circles). The interpolation was performed using a natural cubic spline, which was next sampled to form a 5000-points equally spaced dataset. b) Lomb-Scargle periodogram of the K89 lightcurve: solid black -- original data, dashed blue -- sampled with a time step equal to the mean step of the original data, dotted red -- sampled with a time step short enough to form a 5000-points time-series. The vertical axis is in normalized auxiliary units. A 6.8 d period is visible in the original dataset.
\label{fig1}}
\end{figure}

\citet{klav} obtained 38 data points forming the lightcurve in the $R$ band over more than 50 days. The last point in the series was obtained after an 11 day break, therefore is excluded from this work. The brightness was reported to be constant over a time period of 6 hours at 0.01 mag level. Each night resulted in multiple, independent observations, such that some uncertainties were smaller than 0.01 mag. The data was corrected to mean opposition distances and to zero solar phase angle (for further details, see K89). For the purpose of this work it is useful to resample the unequally spaced data to form an artificial, uniformly spaced lightcurve. In order to do that a cubic spline was formed, which was next sampled with a step equal to the mean of the original dataset to retrieve a set consisting of the same number of points as the original. The statistical properties are gathered in Table~\ref{tbl-1} and the result of this procedure is displayed in Fig.~\ref{fig1}a. This was to verify whether the cubic spline introduces or cancels some structures in the original lightcurve. The inspection of the Lomb-Scargle periodogram \citep{scargle} in Fig.~\ref{fig1}b shows that the periodogram of a cubic spline, sampled with a step equal to the mean separation between datapoints, follows well the periodogram of the original lighturve. However, sampling the cubic spline in order to get 5000 datapoints seems to amplify the frequencies already present in the lightcurve, leaving roughly the same modes. Hence no filtering is applied. For such short lightcurves, producing $\sim$1500 times more datapoints by interpolation must introduce some artificial features, and the aim is to verify whether a nonlinear (possibly, chaotic) dynamical content present in the data might be detected in this way. Next, the cubic spline was sampled with a step chosen so to form a 5000-points dataset. Fig.~\ref{fig1}b presents also the periodogram for this final time-series, that will be used in the estimation of the mLE.

\citet{devyatkin} performed observations broken into two parts by a long interval: from September 1999 to March 2000 and from September to October 2000. Each of the $CBVR$ bands in the respective period will be therefore referred to as $C1$, $C2$, and so on. The intrinsic accuracy was obtained by means of the standard deviation of the average brightness relative to each of the comparison stars in the frame. These values have an average of 0.12 mag in the $C$ band, 0.06 mag for the $B$ band and 0.10 mag for both $V$ and $R$ bands (for further details, see D02). The $B1$ and $V1$ datasets contain only 5 points, and therefore are excluded from the analysis in this work; the $R1$ and $R2$ ones contain 11 points each, though the first observation in $R1$ occured more than a month before the next one, therefore is excluded from the calculations. The number of datapoints in the lightcurves under consideration are gathered in Table~\ref{tbl-1} and range from 10 to 24. The second period is better sampled, with mean spacings from 3.21 to 4.39 days, and standard deviations about 80\% of the mean. The $C1$ sample from the first period is spread over a longer interval, but with a mean spacing equal to 7.86 days (which is roughly 25\% of the minimal Lyapunov time \citep{shevchenkoa}) with a standard deviation of 9.50 days. Each sample was used to form a cubic spline to be spaced with a step equal to the mean of the original dataset. The power spectrum, constructed as in Fig.~\ref{fig1}b, shows this procedure not to be as valid as in the K89 case, although with an undersampled time-series it is difficult to formulate unambiguous conclusions. On the other hand, it is visible in Fig.~\ref{fig1}a that the cubic spline and sampling reproduce the original observations well and do not introduce any auxiliary peaks that would not follow the trend of the inspected time-series, what is also the case in D02 data, therefore we proceed with the analysis. Next, each cubic spline was sampled with a step chosen so to obtain artificial datasets consisting of 5000 points. These, as well as the one obtained from K89, are the subject of the current work.

	\section{Calculation of the mLE}\label{sect3}

		\subsection{Takens reconstruction}\label{subsect3-1}

Before evaluating the mLE, it is insightful to reconstruct the phase-space trajectories {\it via} Takens time delay embedding method \citep{takens}. Having a series of scalar measurements $x(t)$ uniformly distributed at times $t$ one can form an $m$-dimensional location vector using only the values of $x(t)$ at different times given by the classical formula
\begin{equation}
S(t)=[x(t),x(t+\tau),x(t+2\tau),\ldots,x(t+(m-1)\tau)],
\label{eq1}
\end{equation}
where $m$ is the embedding dimension, and $\tau$ is the time delay chosen so that the components of $S(t)$ are independent or uncorrelated. These parameters are estimated {\it via} the False Nearest Neighbor (FNN) algorithm (i.e., embedding dimension $m$) and as the first minimum of the Mutual Information (MI) method (i.e., time delay $\tau$), using the programs described in \citep{kodba,perca,percb,perc06}. These \texttt{.exe} programs (\texttt{fnn}, \texttt{mutual} and others) are available from the website\footnote{\url{http://www.matjazperc.com/ejp/time.html}}. The \texttt{TISEAN} software package \citep{hegger} is also widely used throughout this article\footnote{\url{http://www.mpipks-dresden.mpg.de/~tisean/}}. The MI is an alternative for the commonly used autocorrelation function, where the time delay $\tau$ used to be estimated as the delay at which autocorrelation drops to $1/e$ (to ensure $e$-folding, or, rarely, to 0, in order to form uncorrelated location vectors). This approach will be also explored in the subsequent sections.\footnote{The \texttt{.exe} programs were used as they contain an implementation of the \citet{wolf} algorithm (see next subsections) absent in the \texttt{TISEAN} package, the stationarity test and the program \texttt{mutual} gives the time delay $\tau$ estimated both {\it via} the MI and the autocorrelation function in one run.}

The MI gives the amount of information one can obtain about $x_{t+\tau}$ given $x_t$. In short, the absolute difference between $x_{\rm max}$ and $x_{\rm min}$ is binned into $j$ bins, $j$ being large enough, and for each bin the MI, denoted by $I(\tau)$, is constructed from the probabilities that the variable lies in the $h$-th and $k$-th bins ($P_h$ and $P_k$ respectively and $h,k=1,\ldots,j$) and the joint probability $P_{h,k}$ that $x_t$ and $x_{t+\tau}$ are in bins $h$-th and $k$-th, respectively:
\begin{equation}
I(\tau)=-\sum\limits_{h=1}^{j}\sum\limits_{k=1}^{j}P_{h,k}(\tau)\ln\frac{P_{h,k}(\tau)}{P_h P_k}.
\label{eq2}
\end{equation}

The FNN fraction is calculated under the assumption that the trajectory folds and unforlds smoothly. Roughly speaking, when two initially nearby points diverge under forward iteration (typically not longer than for a time $\tau$), they do so not faster than $R_{\rm tr}\epsilon$, where $\epsilon$ is the initial separation (not greater than the standard deviation of the data)  and $R_{\rm tr}$ is a certain threshold. When this behaviour is changed with the rise of $m$, the points are marked as {\it false nearest neighbors}; the fraction of FNNs decreases with $m$ and reaches a value significantly close to zero (in practical implementations to the threshold $R_{\rm tr}$) for the embedding dimension considered to be a correct estimate of the proper one. This is achieved by calculating the FNN fraction
\begin{equation}
R_i=\frac{|x_{i+m\tau}-x_{t+m\tau}|}{||S(i)-S(t)||}
\label{eq3}
\end{equation}
for nearby points $S(i),\,S(t)$ such that $||S(i)-S(t)||<\epsilon$.

For a detailed description of these algorithms, see \citep{kodba,perca,percb,perc06} and \citep{hegger} for another common implementation. The parameters $m$ and $\tau$ are both required for the mLE calculation.

Fig.~\ref{fig2} presents the phase-space reconstruction for the considered time-series, using the MI. The FNN fraction indicated that for most of the embeddings $m=2$ is sufficient, being \mbox{$m=3$} only for K89 and $C1$ datasets (note these are the best sampled long-term observations obtained). Yet, for consistency with the rest of the plots and due to a low dimension of the reconstructed phase-space trajectories (see next subsection) all trajectories shown are embedded in an $m=2$ space so that the figures displayed are more perspicuous. The embeddings using time delays from the autocorrelation function were also performed and are displayed in the Appendix. The phase-space reconstructions seem to posses the same topological structure, which corroborates that the sampling of a cubic spline, as described in the previous section, allows to reveal dynamics that they stem from.

		\subsection{Correlation dimension}\label{subsect3-2}

The correlation dimension $d_C$ is a measure of how much space is covered by a set. For usual 1D, 2D or 3D cases the $d_C$ is equal to 1, 2 and 3, respectively. However, there are sets called fractals, which posses a fractional correlation dimension. For dissipative dynamical systems, e.g. the Lorenz system \citep{lorenz}, chaotic motion manifests itself through a limitting trajectory called the strange attractor, due to its fractal properties. Herein we examine the fractal dimension of the reconstructed phase-space trajectories in order to further constrain the proper embedding dimension. For Hamiltonian systems, i.e. volume preserving, this is not an indicator of chaoticity \citep{greiner}.

The correlation dimension is defined as
\begin{equation}
d_C=\lim_{r\rightarrow 0}\frac{\ln C(r)}{\ln r},
\label{eq4}
\end{equation}
with the estimate for the correlation function $C(r)$ as
\begin{equation}
C(r)=\lim_{N\rightarrow\infty}\Bigg[\frac{2}{N(N-1)}\sum_{i=1}^N\sum_{j=i+1}^N H(r-||x_i-x_j||)\Bigg],
\label{eq5}
\end{equation}
where the Heaviside step function $H$ adds to $C(r)$ only points $x_i$ in a distance smaller than $r$ from $x_j$ and {\it vice versa}. The limit in Eq.~(\ref{eq4}) is attained by using multiple values of $r$ and fitting a straight line to the linear part of the obtained dependencies. The correlation dimension is estimated as the slope of the linear regression. The calculations were performed with a \texttt{MATLAB} code with the Theiler window equal to zero \citetext{Seymour, priv.\ comm.} and the error calculations are described in \citep{seymour}. The results in a graphical form are shown in Fig.~\ref{fig3}; the numerical values are gathered in Table~\ref{tbl-2}.

Following the reasoning of \citet{grassberger,seymour} one could infer, due to the fractal dimensions not being much greater than unity, that there would be a total of two dynamical variables governing Hyperion's rotation at the time of observations. However, it is well known that chaos can not occur in a two dimensional continuous dynamical system (see Poincar\'e-Bendixson theorem, e.g. \citep{alli}); therefore, there must be at least three variables to consider the rotation being chaotic. This requirement is met for example in the simplified model (with 1.5 dof) in which the axis of rotation is fixed and perpendicular to the orbit plane \citep{wisdom,celletti}. This leads to the suspicion that the underlying dynamics are not governed by the chaotic zone, i.e. Hyperion remained in a regular (quasi-periodic) state, possibly influenced by noise. On the other hand, the datasets with $m=2$ are undersampled and the chaotic behaviour may not be visible. Datasets with $m=3$ consist of slightly more observations, therefore the FNN algorithm may have caught the occurence of nonlinear phenomena. Still, the lengths of the time series are much smaller than required for an unambiguous analysis \citep{grassberger}, and the very low correlation dimensions attained are a sign of this. The sampling might also ruled out some nonlinear (chaotic) features, however, as the original data is unevenly sampled, this is hardly to be avoided. Finally, it was shown \citep{ruelle} that dimension estimates that are not below $2\log N$ are not reliable. Herein, values near unity are obtained, which fall below this limit, but due to shortness of the datasets one cannot really infer any reasonable estimate for the correlation dimension, especially that Hyperion's dynamics in fact is located in a six-dimensional phase space \citep{wisdom,klav,devyatkin,shevchenkoa,shevchenkob,kouprianov} due to being well described by a full set of Euler equations. For possible rotational--lightcurve models see \citep{hicks}.

		\subsection{Maximal Lyapunov Exponent}\label{subsect3-3}

The mLE, denoted $\lambda$ in general, was calculated using two distinct algorithms: the \citet{wolf} and \citet{kantz94} methods, incorporated in the programs \texttt{lyapmax} and \texttt{lyapmaxk} \citep{kodba,perca,percb,perc06}, and \texttt{lyap\_k} from the \texttt{TISEAN} package. Herein the results of those investigations are presented.

The time delay $\tau$ was calculated in Section~\ref{subsect3-1}. The embedding dimension $m$ was not set according to the FNN results, but was varied from $m=2$ to $m=10$ and first the mLE was computed using the approach of \citeauthor{wolf} The algorithm finds a nearest neighbor to an initial point and evolves them both until the separation becomes too big; next, the distance is being rescaled in order to stay in the small-scale structure, and this repeats to the end of the time-series. Then, the average of the logarithms of the displacement ratios is the estimate of the mLE. All of the numerical results are gathered in Table~\ref{tbl-3}, while Fig.~\ref{fig4} displays the mean values of each dataset. Due to the concluding remark from the previous subsection, $\bar{\lambda}_{\rm max}$ is computed including and excluding the $m=2$ values. The exclusion leads to lessening the standard deviation of the $\bar{\lambda}_{\rm max}$ corresponding to $C1$ data and changing the sign of K89 and $R1$ datasets' mLE to negative with significant diminishing their standard deviation. Because, as mentioned, chaos cannot be present in a two-dimensional continuous phase-space, the results from Fig.~\ref{fig4}b seem to be more realistic. The FNN convergence to $m=2$, as shown in Table~\ref{tbl-3}, corresponding to positive Lyapunov exponents, must be an artifact due to either numerical limitations of the algorithm, or to undersampling of the lightcurves. Analogous results using the autocoreelation function are shown in the Appendix. The main drawbacks of the \citeauthor{wolf} method are that {\it i)} it fails to take advantage of all available data as it focuses on one fiducial trajectory \citep{rosenstein} and {\it ii)} it does not test the presence of exponential divergence (a behaviour underlying chaotic dynamics) but assumes it explicitly {\it ad hoc}, what may lead to spurious results \citep{kantz04}.

The Kantz method differs from the previous in that it takes several points in a neighborhood of some particular point $x_i$. Next, one computes the average distance of all obtained trajectories to the reference, $i$-th one, as a dependence of the relative time $n$ (incorporated to the $k$-th subscript as follows: $x_{k+(m-1)\tau+n}$). The average $S(n)$ of logarithms of these distances is plotted as a function of $n$ and the slope of the linear part is the mLE (see \citep{perca} for further details). In the case of chaos, three regions should be distinct: a steep increase for small $n$, a linear part and a plateau \citep{seymour}.

The results shown in Fig.~\ref{fig5} clearly show no linear part in the plots, therefore one could conclude that there is no positive mLE, indicating lack of chaotic rotation in the examined lightcurves. Yet, according to previous research (ground-based observations under investigation herein as well as based on Voyager~2 and Cassini images), a chaotic rotational state is undoubtful. On the other hand, Fig.~\ref{fig6} displays the $S(n)$ relation obtained for K89 and $C1$ data using the autocorrelation to estimate the time delay. Surprisingly, only one of almost 30 dependencies displayed for each shows a linear part. Moreover, that is clearly a spurious detection, as the thick red lines in Fig.~\ref{fig6} are related to embedding dimension $m=2$, in which chaos can not be present. However, the methods described require the data to fulfill the {\it stationarity} assumption, i.e. that the statistical properties of a time-series (e.g., mean and standard deviation) are constant in time. We therefore perform a test in a following way \citep{perc06}.

		\subsection{Stationarity test}\label{subsect3-4}

Let us consider a point $p(t)$ as an event and find all {\it similar} events, i.e. those points $p(i)$ that lie not further than $\epsilon$ from $p(t)$. We average all values of $x_i$ and call this a prediction of a future observation based on the value of $x_t$. The key now is to use cross-prediction, i.e. to partition the whole dataset into non-overlapping segments and use the $j$-th segment to make a prediction of a $k$-th segment. We quantify the correctness by calculating the error $\delta_{jk}$ {\it via} square root of mean square deviations from the mean in segment $k$ and repeat this for all $j$ and $k$:
\begin{equation}
\delta_{jk}=\sqrt{\frac{1}{N}\sum\limits_{k=1}^N (\tilde{x}_k-x_k)^2},
\label{eq6}
\end{equation}
where $\tilde{x}$ is the prediction and $x$ is the true value in a $k$-th segment. If $\delta_{jk}$ is significantly larger than the average, then either the dynamics are not conserved from one segment to another, or the data is undersampled. Both cases yield a conclusion that the data is non-stationary.

The program \texttt{stationarity} was applied to sampled lightcurves under consideration and the results are gathered in Fig.~\ref{fig7}.

The immediate denouement is that the lightcurves from K89 and D02 are too short, undersampled, or both. Therefore, it is justified to ask a question: how long and how dense should photometric observations be in order to reveal a positive mLE in a lightcurve?

	\section{Numerical experiments}\label{sect4}

To answer the last question, we examine simulated lightcurves of Hyperion for chaotic and regular solutions of the Euler system of equations. These lightcurves, as well as the LEs spectra, were obtained from \citetext{Mel'nikov, priv.\ comm.} and were computed using a procedure described in D02. The algorithm gives Hyperion's stellar magnitude $m$ in time (JD), corrected to zero solar phase angle and mean opposition magnitude, as well as the time evolution of the Euler angles ($\theta,\,\varphi,\,\psi$) and the corresponding angular velocities ($\omega_1,\,\omega_2,\,\omega_3$). Table~\ref{tbl-4} gathers parameters and initial conditions necessary to run simulations of lightcurves as described in D02 and obtained from \citetext{Mel'nikov, priv.\ comm.}. Full spectra of LEs were calculated using the HQRB method \citep{bremen} realized as a software complex in \citep{shevchenkob,kouprianov}. The described data are shown graphically in Fig.~\ref{fig8} together with the output of the stationarity test. The system is Hamiltonian, therefore the six LEs are paired so that $\lambda_i+\lambda_j=0$ and the plots show only three positive LEs. The Lyapunov time $T_L$ for the chaotic solution is equal to 44 d.

Since the lightcurves have a constant time step $\Delta t=0.1\,{\rm d}$ (hence consist of $\sim 10^4$ data points), in order to produce time-series more astronomically realistic only three first points during each day were left and averaged (see Fig.~\ref{fig8}). Then a cubic spline and sampling were performed to produce datasets consisting of 5000 points. From these sampled lightcurves, intervals of lengths: 2 months, 6 months and 1 year were chosen randomly; each had ten realisations both for the chaotic and regular solution. The whole 3 year lightcurves were taken as single realisations. Next, the routines \texttt{false\_nearest} and \texttt{autocor} from the \texttt{TISEAN} package were applied for obtaining the time delay $\tau$ and \texttt{lyap\_k} for embedding dimension from 2--10 to extract the mLE. In the same way time dependencies of dynamical variables $(\theta,\,\varphi,\,\psi,\,\omega_1,\,\omega_2,\,\omega_3)$ were investigated. It may be surprising that there are clearly linear parts in the stretching factor $S(n)$ plots in the regular case, however, a closer look at the stationarity tests show that the variables showing false chaotic behaviour are non-stationary, what may influence the divergence exhibited by the $S(n)$ dependence. Therefore, we do not need to be worried by this confusing result, yet it is worth noting that in the case of real astronomical observations, if the stationarity test is omitted, one can easily find chaotic phenomena where they are in fact absent. To illustrate this statement, Fig.~\ref{fig9} displays time evolutions of variables having a linear part in the $S(n)$ plot and the corresponding stationarity tests. For a chaotic solution, we conclude that all time-series are stationary enough to proceed with the investigation.

For all subsets the time delay was determined using the autocorrelation function, the $S(n)$ plots were computed and inspected for presence of linear regions, indicator of chaotic rotation, potentially visible {\it via} photometric observations. For comparison, corresponding subsets from the original simulated lightcurves (i.e., having a time step of 0.1 d) were examined to check what information about the mLE is lost due to sampling effects. For two-month intervals no linear part is visible in the stretching factor. In half-year subsets some traces of linearity are noticeable, but they are not unequivocal enough to claim a detection. On the other hand, in the original datasets more unambiguous chaotic behaviour is present. In one-year segments the chaoticity occurs more frequently and is supported by its presence in the original lightcurve's $S(n)$ plots. The whole, three-year long, lightcurve yields a confirmation of chaotic rotation. In Fig.~\ref{fig10} some representatives for each subset length are displayed. The lightcurve simulated for regular rotation gives no linear regions in any of the $S(n)$ plots, as expected, but due to non-stationarity of the sampled data, in some of the stretching factors a linear rise is more evident than for the simulated chaotic case. This proves that stationarity is a necessary condition for a dependable detection of chaotic phenomena.

	\section{Discussion and conlusions}\label{sect5}

The aim of this paper was to verify whether it is possible to infer the value of the mLE from photometric observations of Hyperion. Firstly, existing datasets (K89 and D02) were investigated using Takens phase-space reconstruction and its correlation dimension, stationarity tests and finally the mLE was estimated using two algorithms: \citet{wolf} and using stretching factors \citep{kantz94}. \citeauthor{wolf} method yielded a positive detection for K89 observations, nevertheless the $S(n)$ plots showed no linear region, implying lack of chaoticity. As elaborated, the \citeauthor{wolf} method is likely to yield spurious detections, especially for short datasets. We therefore conclude that existing datasets are too short and undersampled to detect chaotic rotation using the mLE.

In order to list conditions allowing to obtain the mLE from potential ground-based observations, simulated lightcurves spanning 3 years were examined. As suggested from previous considerations, two-month long subsamples appeared to be too short to yield a sign of chaos. Half-year data retained a clearly visible linear rise in some of the $S(n)$ plots. On the other hand, to make these samples more atronomically realistic, only magnitudes spanning $\approx$ 7.5 h each day (more precisely, night observations) were left and averaged. This led to a conclusion that only one year subsets are long enough to reveal the presence of chaos in the stretching factor, but only in favorable conditions.

Additionally, a false detection of chaos was observed in the case of lightcurves based on regular solutions of the Euler equations. To explain this extraordinary behaviour a careful inspection of the stationarity test outputs was conducted. It was found that the time-series underlying the simulated lightcurve are non-stationary (as well as the lightcurve itself) which violates the assumptions underlying the mLE calculation algorithm. Therefore, obtaining a positive mLE is by itself not sufficient to claim detection. A necessary stationarity condition must be also fulfilled.

Based on computations described herein we assert than to reliably estimate presence of chaos in ground-based photometric observations of Hyperion {\it via} mLE, these observations should be performed over a time period of at least one year. A way to shorten this period is to obtain well-sampled photometry, e.g. by observing with 2--3 telescopes spread over the world. As was noted, in case of data points distributed uniformly with a time step equal to 0.1 d, timespan may be shortened to half year. We remind that the resulting time-series should be stationary. However, even with long-term observations, it might happen that Hyperion will temporarily remain in a dynamical state that will not allow to make any conclusive claims about its rotation.

\acknowledgments

The author is grateful to Andrew Seymour for discussions and sharing the \texttt{MATLAB} code, and especially to Alexander Mel'nikov for helpful discussions and providing useful information as well as the simulated lighcurves.

\appendix

	\section{Appendix}\label{app}

\begin{figure}
\plotone{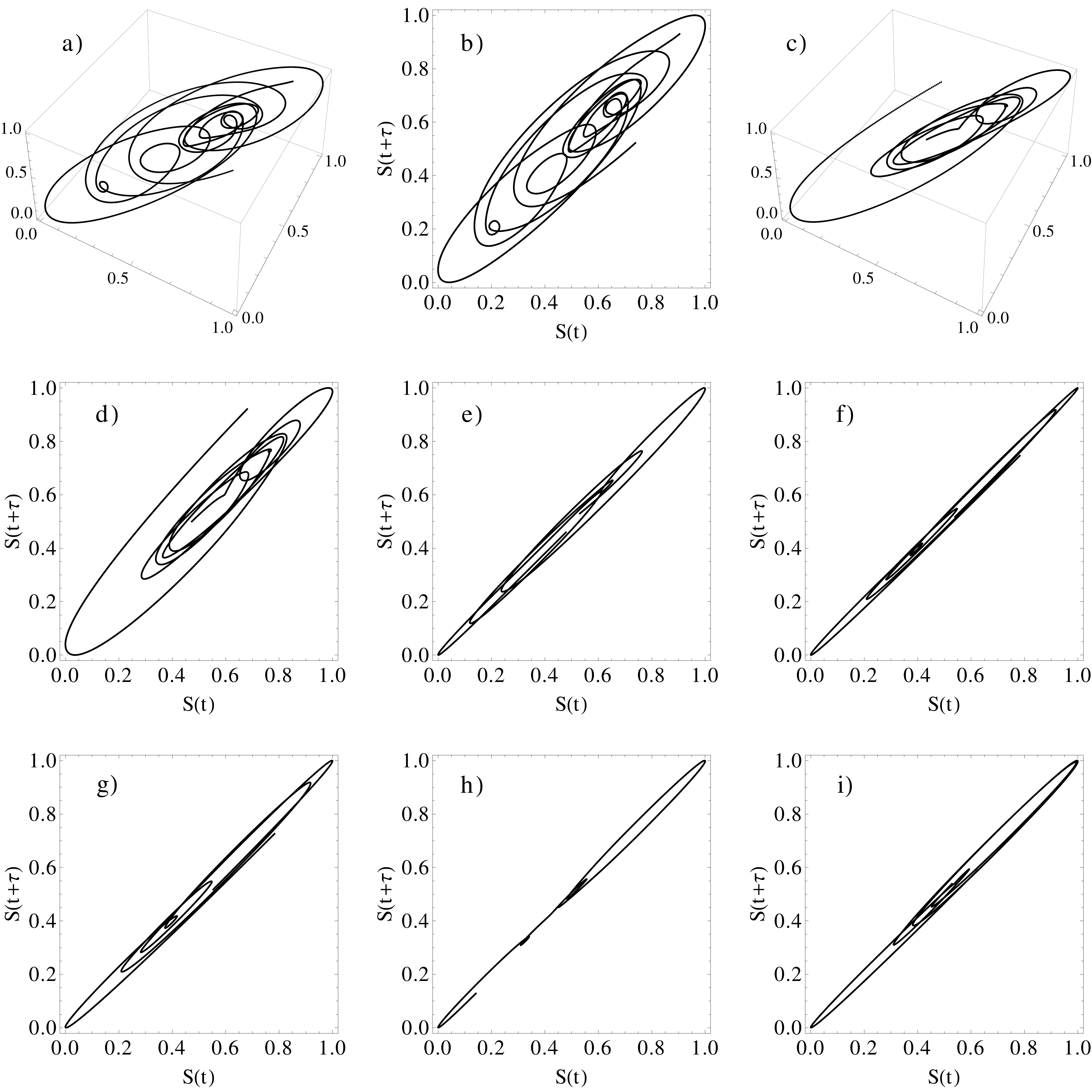}
\caption{Phase-space trajectories reconstructed using the Takens delay time method and normalized to a unit box. All embeddings appear to posses the same topology, indicating each trajectory stems from the same underlying dynamics. The corresponding datasets are: a) K89 in 3D b) K89 in 2D, c) $C1$ in 3D, d) $C1$ in 2D, e) $C2$, f) $R1$, g) $R2$, h) $B2$, i) $V2$. The delay $\tau$ is different for each embedding and estimated using the MI approach. Note that despite all reconstructions posses the same topology, e) -- i) look like being of a purely regular time-series. This may be due to the undersampling of their corresponding lightcurves.
\label{fig2}}
\end{figure}

\begin{figure}
\plotone{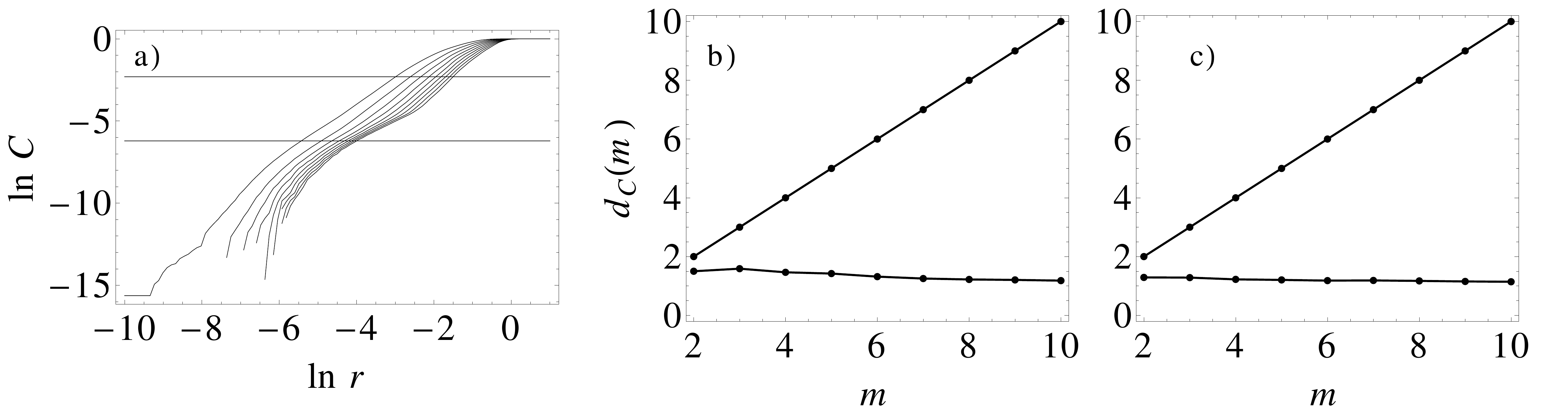}
\caption{Estimates of the correlation dimension of the reconstructed phase-space trajectories. a) The $\ln C(r)$ vs. $\ln r$ plot of the K89 dataset. The lines correspond to $m=2$ (left) up to $m=10$ (right). The horizontal lines mark the cutoff values of the $C(r)$: the lower one is at the level of $C(r)=10/N$, $N$ being the total length of the time series, and the higher one is for $C(r)=0.1$. b) The line with a slope of 45 degrees shows the correlation dimension for a purely random time series at given $m$. The flat line shows the correlation dimension for the actual time series (K89). c) The same as b), but for $C1$ data. Other results for D02 are not shown because they are visually indistinguishable from a plot for $C1$.
\label{fig3}}
\end{figure}

Embedding delays $\tau$ for phase-space reconstruction {\it via} Takens method were calculated as the delay at which the autocorrelation function drops to 1/e, leading to values being an order of magnitude greater than those from the MI algorithm. The reconstructions are displayed in Fig.~\ref{fig11}, while correlation dimensions of these embeddings are uninsightful and therefore not presented herein. Note that all trajectories are characterized by correlation dimension not much greater than unity; furthermore, the 3D embeddings in both cases show no intersections, which is a premise that $m=3$ is sufficient. Figs.~\ref{fig2}~and~\ref{fig11} reveal intersections in 2D plots, what is naturally a projection effect.

\begin{figure}
\plotone{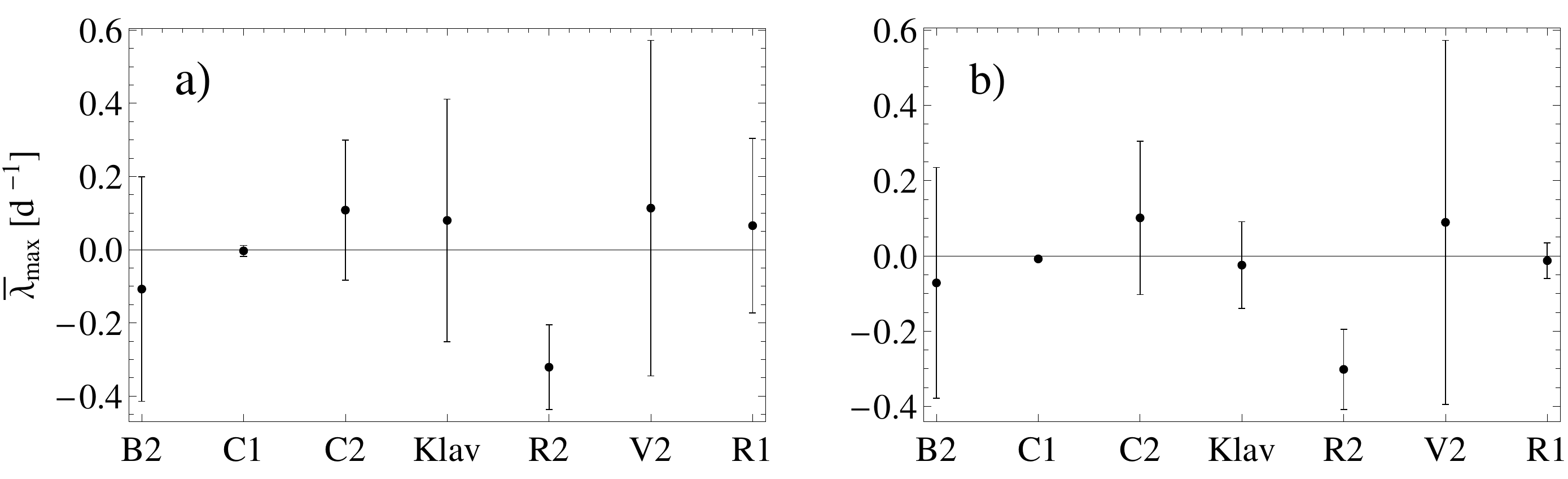}
\caption{The mean values of the mLE $\lambda_{{\rm max}}$ of \citeauthor{wolf} algorithm, calculated over $m$ a) from 2 to 10 and b) from 3 to 10, using the MI to estimate the time delay; the error bars mark the standard deviation of the mean.
\label{fig4}}
\end{figure}

\begin{figure}
\plotone{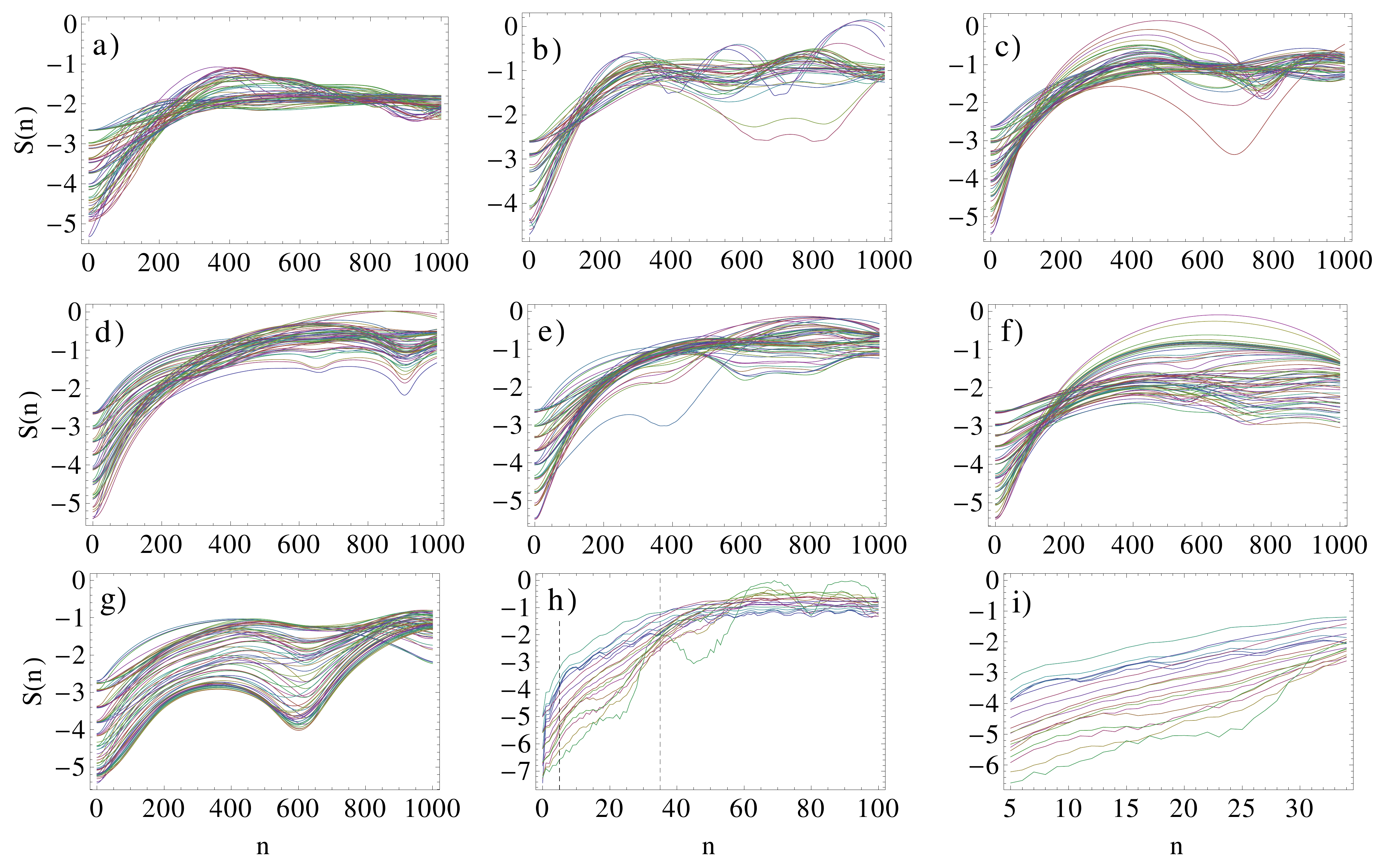}
\caption{Calculation of the mLE using the Kantz algorithm for a) K89, b) $C1$, c) $C2$, d) $R1$, e) $R2$, f) $B2$, g) $V2$ datasets. None of these plots exhibit a linear part, contrary to an evident case of h) Lorenz system. Beams of curves starting at different levels at $n=0$ refer to different neighborhood sizes. Each curve in a beam is for a different $m$ from 2 to 10, starting at the lowest curve. For comparison, h) displays a plot for the $x$-component of length $t_{\rm max}=500$ of the Lorenz system; the vertical dashed lines mark the section of a linear trend and i) shows its magnification. The mLE, which is ten times the slope (due to sampling with $\Delta t=0.1$), is equal to $0.90\pm 0.04$ with $R^2=0.95$, while calculation using the system's differential equations yield 0.90, which is an excellent agreement.
\label{fig5}}
\end{figure}

\begin{figure}
\plotone{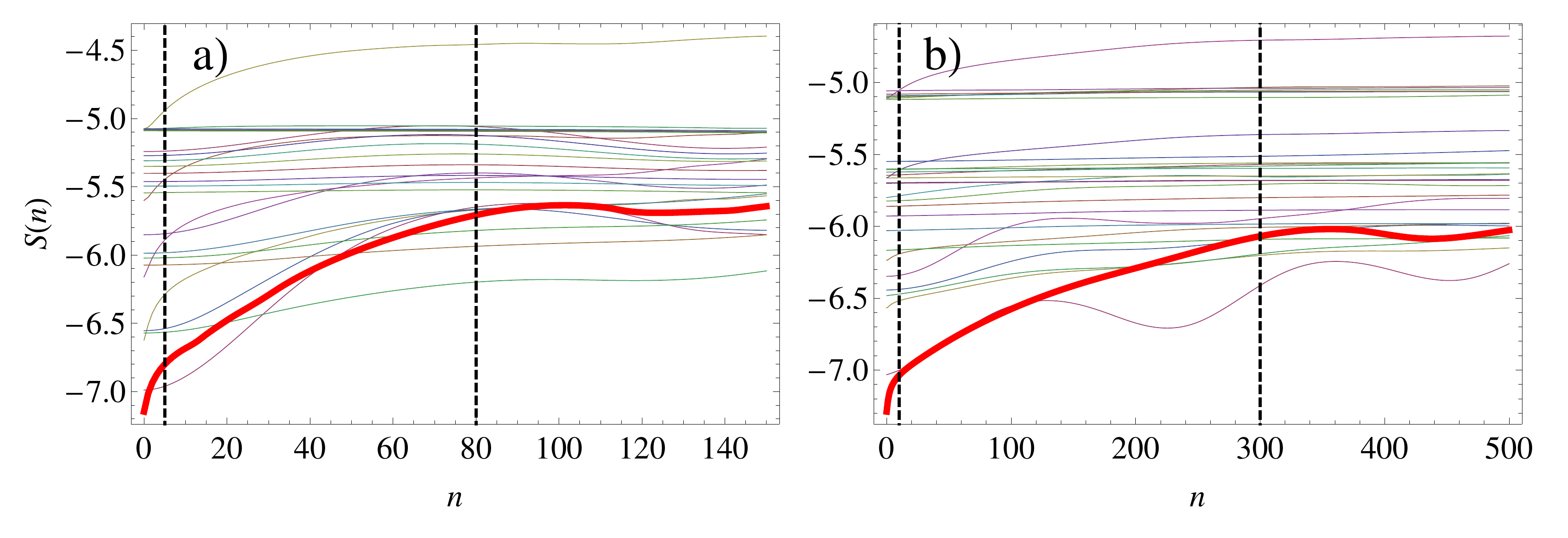}
\caption{Calculation of the mLE using the Kantz algorithm and the autocorrelation function for a) K89 data and b) $C1$ data. Vertical dashed lines mark the linear region of the thick, red lines. The slopes are equal to $0.0145\pm 0.0003$, $R^2=0.99993$, and $0.00310\pm0.00003$, $R^2=0.99995$, respectively.
\label{fig6}}
\end{figure}

\begin{figure}
\plotone{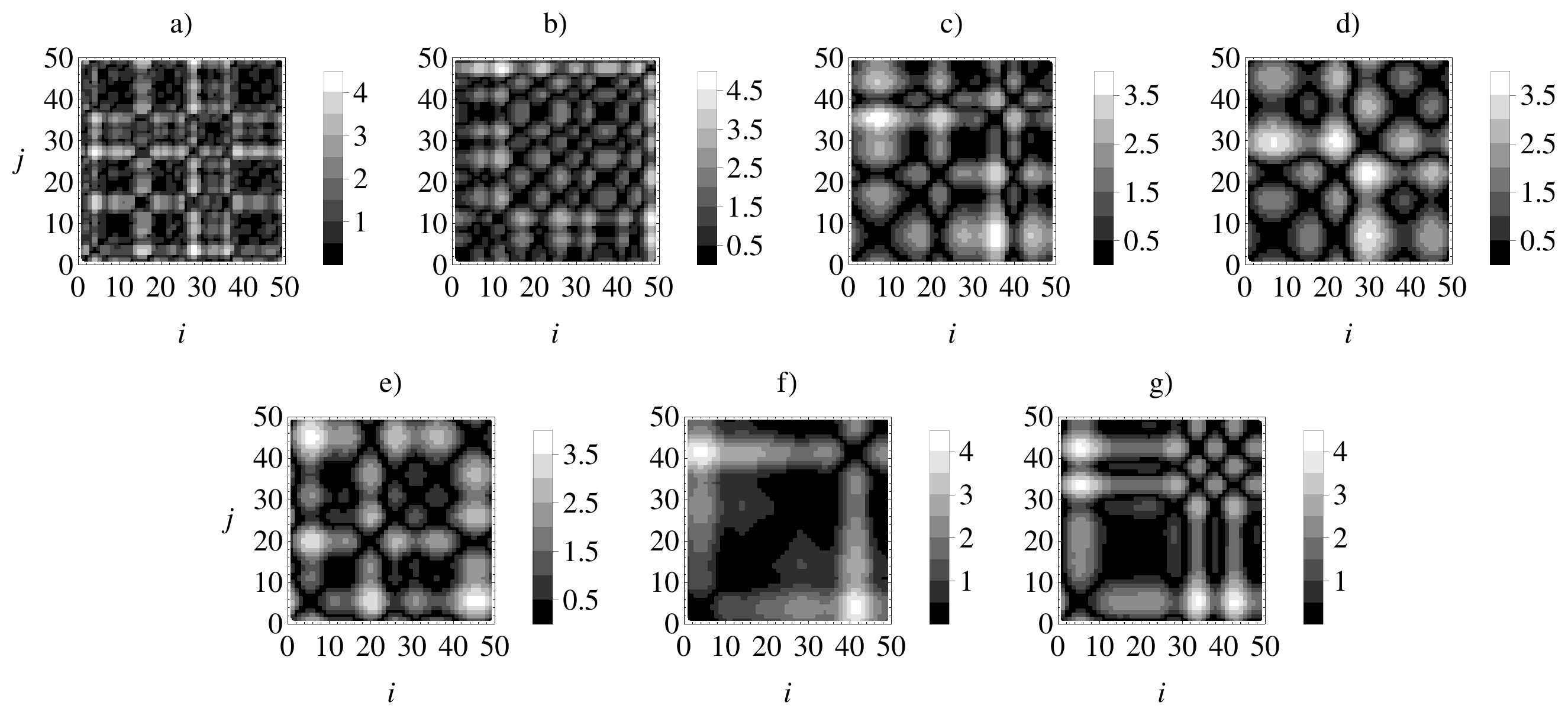}
\caption{Stationarity tests for a) K89, b) $C1$, c) $C2$, d) $R1$, e) $R2$, f) $B2$, g) $V2$. All sampled lightcurves are non-stationary, the most uniform being $C1$ data, yet having high prediction error fluctuations.
\label{fig7}}
\end{figure}

\begin{figure}
\plotone{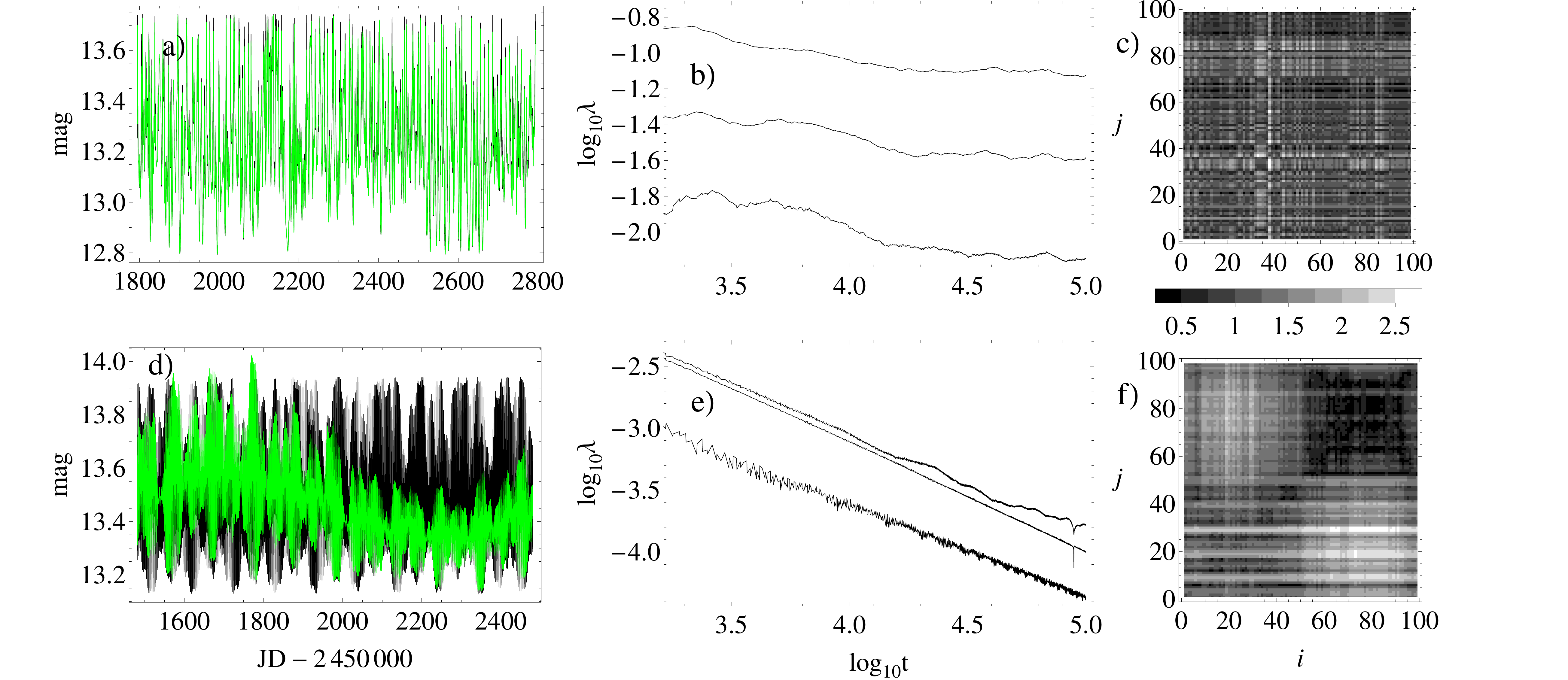}
\caption{a) The simulated chaotic lightcurve (black) and the sampled one (green); b) the convergence of the LEs and c) the stationarity test of the sampled (green) lightcurve. d)--f) the same as a)--c), but for the regular solution, which is clearly non-stationary.
\label{fig8}}
\end{figure}

\begin{figure}
\plotone{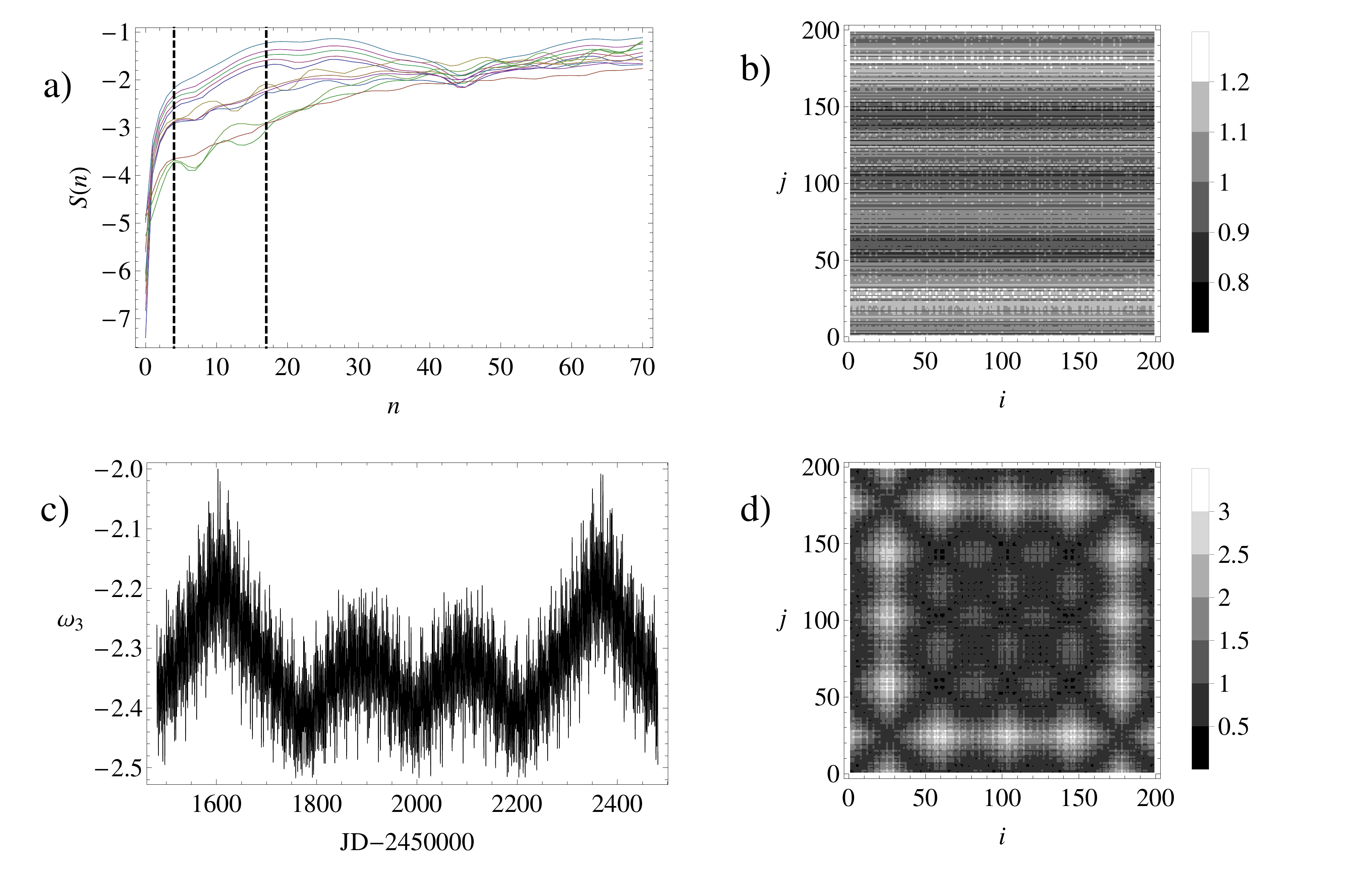}
\caption{a) The stretching factor for the $\psi$ component for regular rotation; vertical dashed lines mark the linear region; b) the corresponding stationarity test. c) Time evolution of the $\omega_3$ component; the trend is clearly visible; d) the corresponding stationarity test which clearly displays the overall periodicity of the time-series. The stretching factor plot for this variable also unveils a linear part.
\label{fig9}}
\end{figure}

\begin{figure}
\plotone{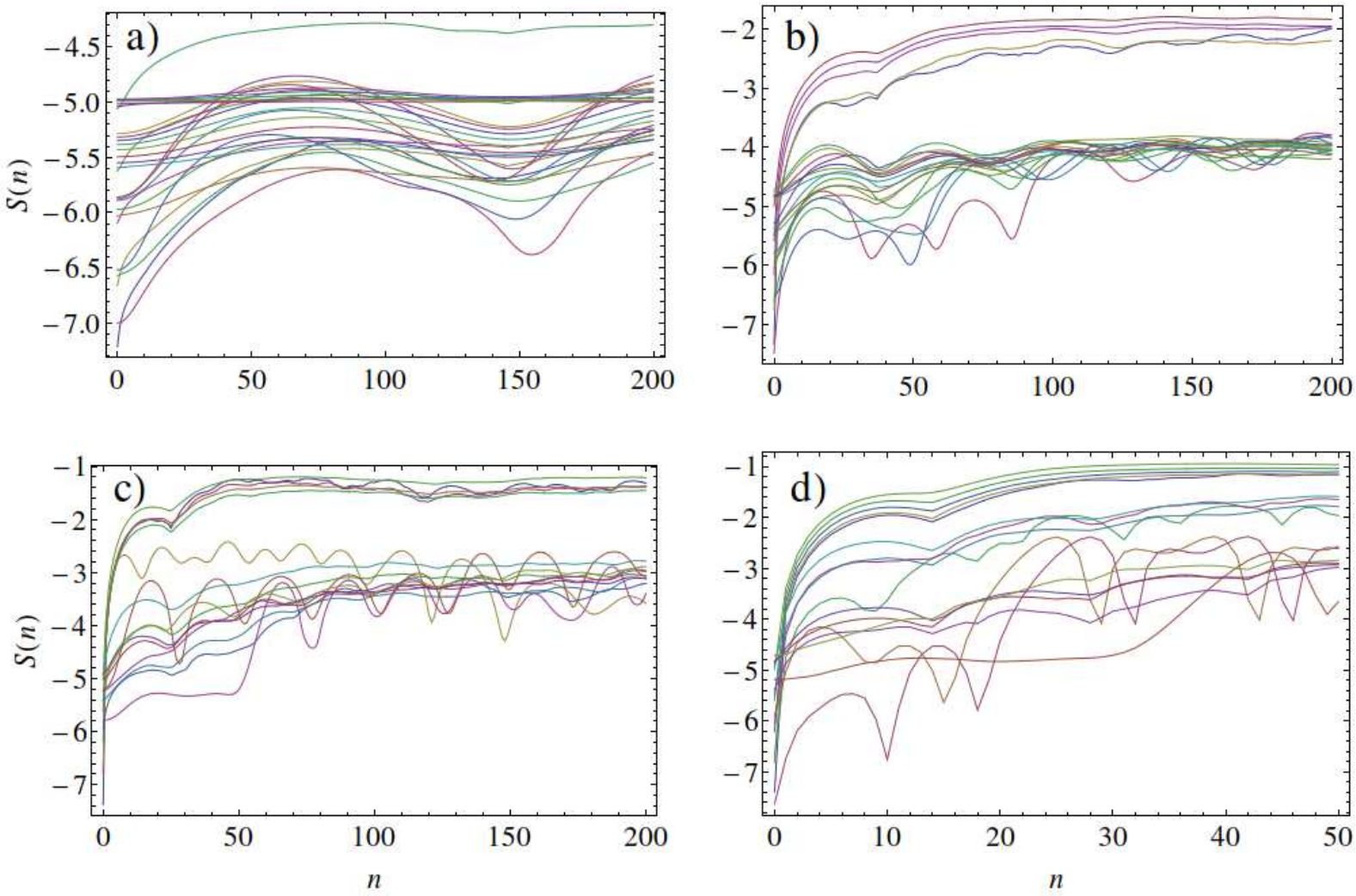}
\caption{Stretching factors for subsets of length a) 2 months, b) 6 months, c) 1 year and d) 3 years. Starting from c) a linear region is visible for higher embedding dimension (lower curves).
\label{fig10}}
\end{figure}

\begin{figure}
\plotone{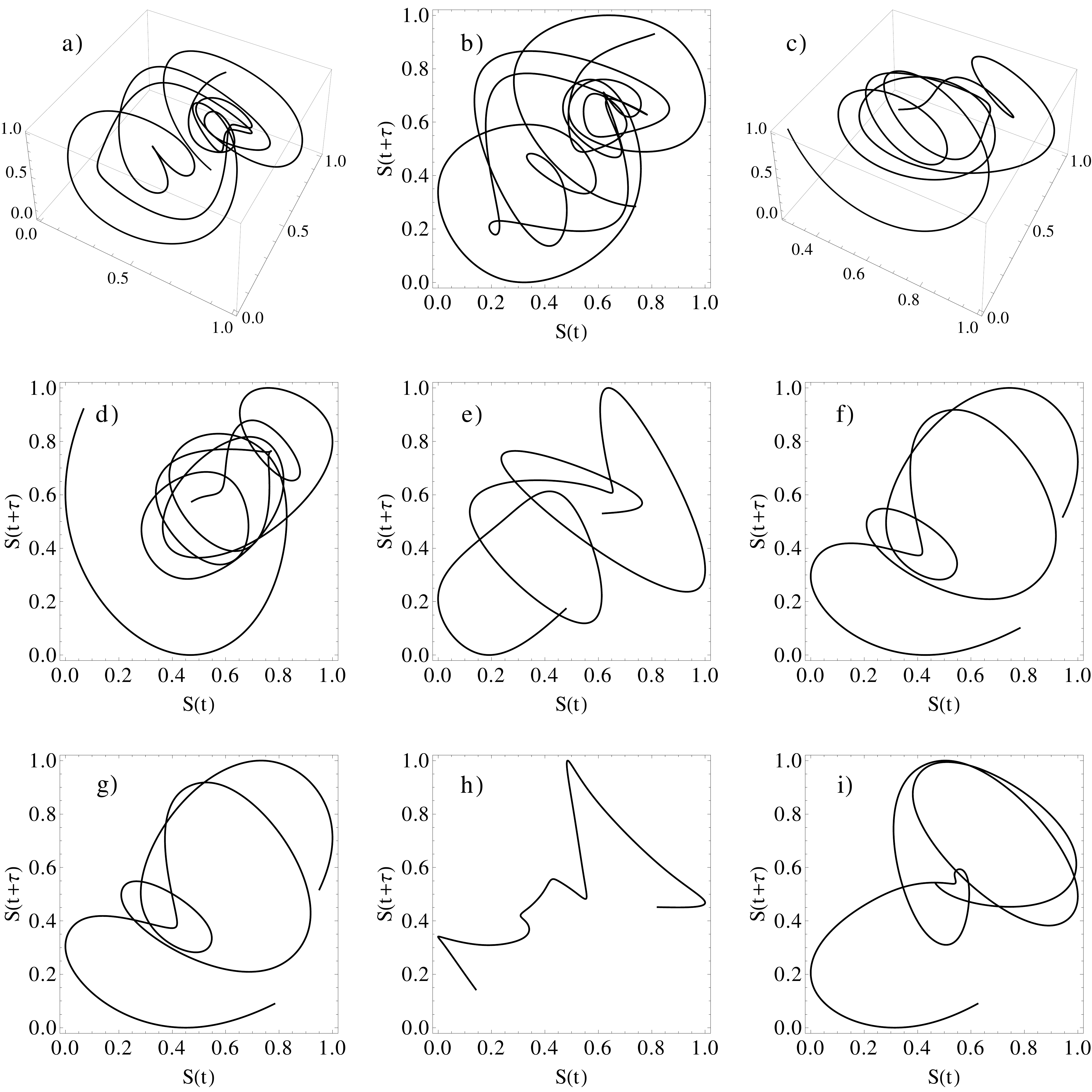}
\caption{Phase-space trajectories reconstructed using the Takens delay time method and normalized to a unit box. All embeddings appear to posses roughly the same topology, indicating each trajectory stems from the same underlying dynamics. The corresponding datasets are: a) K89 in 3D b) K89 in 2D, c) $C1$ in 3D d) $C1$ in 2D, e) $C2$, f) $R1$, g) $R2$, h) $B2$, i) $V2$. The delay $\tau$ is different for each embedding and estimated using the autocorrelation function.
\label{fig11}}
\end{figure}

\begin{figure}
\plotone{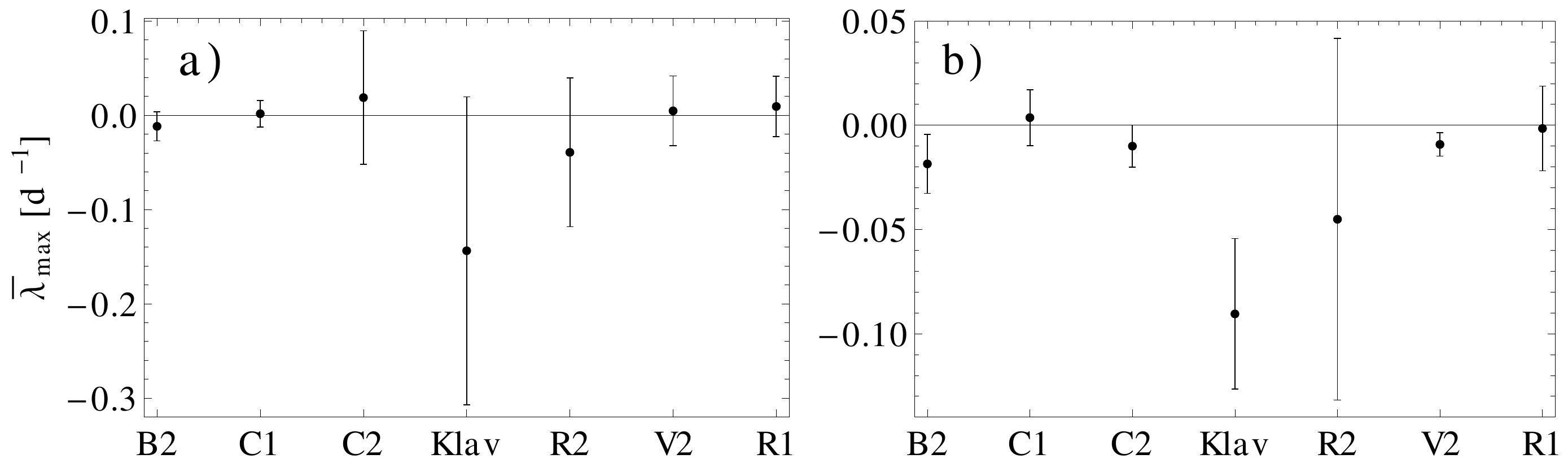}
\caption{The mean values of the $\lambda_{{\rm max}}$ of \citet{wolf} algorithm, calculated over $m$ a) from 2 to 10 and b) from 3 to 10, using the autocorrelation function to estimate the time delay; the error bars mark the standard deviation of the mean. All values obtained for K89 data are negative, while almost all the positive cases are unambiguous due to oscillating behaviour of the mLE.
\label{fig12}}
\end{figure}

Average values of mLEs computed with the \citeauthor{wolf} method are presented in Fig.~\ref{fig12} and in many instances give results contrary to those from Fig.~\ref{fig4} obtained using the MI for estimating time delays. Moreover, the mLE convergence plots frequently shows behaviour oscillating around zero, preventing to see a tendency for a certain sign. As argued in the main text, even when the \citeauthor{wolf} method shows convergence to a positive mLE, this is an ambiguous detection, which may be spurious due to an assumption of exponential divergence of initially nearby trajectories, not necessarily to be met in actual time-series.

\clearpage

\begin{deluxetable}{ccccc}
\tablewidth{0pt}
\tablecaption{Statistical properties of the lightcurves.\label{tbl-1}}
\tablehead{
\colhead{Dataset} & \colhead{No. of obs.} & \colhead{Mean [d]\tablenotemark{a}} & \colhead{Std. [d]\tablenotemark{a}} & \colhead{Median [d]\tablenotemark{a}}}
\startdata
K89 & 37\tablenotemark{b} & 1.47 & 1.00 & 1.00\\
$C1$ & 24 & 7.86 & 9.50 & 5.08\\
$C2$ & 15 & 3.21 & 2.81 & 1.99\\
$R1$ & 10 & 8.22 & 6.52 & 8.03\\
$R2$ & 11 & 4.40 & 3.73 & 3.53\\
$B2$ & 12 & 3.99 & 2.79 & 3.08\\
$V2$ & 13 & 3.66 & 2.95 & 3.05\\
\enddata
\tablenotetext{a}{The 3rd, 4th and 5th columns refer to the spacings between consecutive observations.}
\tablenotetext{b}{In K89, 38 datapoints are listed, but one of them is separated from the others by an 11-day gap and is therefore excluded from the analysis herein.}
\end{deluxetable}

\clearpage

\begin{deluxetable}{ccc}
\tablewidth{0pt}
\tablecaption{Correlation dimensions of the reconstructed phase-space trajectories.\label{tbl-2}}
\tablehead{
\colhead{Dataset} & \colhead{\begin{tabular}[c]{@{}c@{}}Mean correlation\\dimension\end{tabular}} & \colhead{\begin{tabular}[c]{@{}c@{}}Standard\\deviation\end{tabular}}}
\startdata
K89 & 1.31 & 0.13\\
$C1$ & 1.18 & 0.03\\
$C2$ & 1.20 & 0.06\\
$R1$ & 1.25 & 0.03\\
$R2$ & 1.26 & 0.04\\
$B2$ & 1.128 & 0.005\\
$V2$ & 1.05 & 0.02\\
\enddata
\end{deluxetable}

\clearpage

\begin{deluxetable}{cccccccc}
\tablewidth{0pt}
\tablecaption{\citeauthor{wolf} mLE for embedding dimensions from 2 to 10. Values corresponding to $m$ obtained from the FNN algorithm are written in bold. Units are in d$^{-1}$.\label{tbl-3}}
\tablehead{
\colhead{Dataset} & \colhead{K89} & \colhead{$C1$} & \colhead{$C2$} & \colhead{$R1$} & \colhead{$R2$} & \colhead{$B2$} & \colhead{$V2$}}
\startdata
$m=2$ & .915057 & .0334368 & {\bf .163843} &{\bf .691042} & {\bf --.477822} & {\bf --.395662} & {\bf .307586}\\
$m=3$ & {\bf .0527462} & {\bf --.0141602} & .563768 & .062527 & --.537677 & .301896 & 1.0543\\
$m=4$ & --.0877841 & --.00857353 & --.155273 & .051306& --.169326 & --.7375 & --.715765\\
$m=5$ & --.223674 & --.0113392 & .119432 & --.0149389 & --.247616 & .161246 & --.0898974\\
$m=6$ & --.152557 & --.012639 & .0407382 & --.0675292 & --.265254 & .0537108 & --.00898974\\
$m=7$ & .0130682 & --.0129986 & .0656708 & --.0609723 & --.273675 & --.0337969 & .0475659\\
$m=8$ & .0730114 & --.00265503 & .0603281 & --.0135194 & --.298596 & --.0815904 & .0957009\\
$m=9$ & .0805871 & --.00212955 & .0564324 & --.0207522 & --.329889 & --.108901 & .124832\\
$m=10$ & .0469697 & --.000608444 & .055208 & --.0385302 & --.291996 & --.130522 & .203805\\
\enddata
\end{deluxetable}

\clearpage

\begin{deluxetable}{ccc}
\tablewidth{0pt}
\tablecaption{Initial conditions for computing time evolution of dynamical variables used for obtaining the simulated lightcurves.\label{tbl-4}}
\tablehead{
\colhead{} & \colhead{Chaotic} & \colhead{Regular}}
\startdata
$A$ &  0.5662447 & 0.5752270\\
$B$ &  0.6989932 & 0.7008151\\
$C$ &  1.0000000 & 1.0000000\\
$\theta$ &  2.0471549 & 0.2206428\\
$\varphi$ &  0.4684503 & 0.0000000\\
$\psi$ &  3.0482151 & 3.0801234\\
d$\theta$ &  1.1232298 & 2.2203451\\
d$\varphi$ &  0.0622591 & -0.7384962\\
d$\psi$ &  0.1536737 & 1.8101437\\
$H$ &  12.7944259 & 12.7944259\\
$G$ &  -0.9489654 & -0.9489654\\
JD &  2451794.5 & 2451481.3\\
\enddata
\end{deluxetable}

\end{document}